\documentclass[12pt]{article}
\usepackage{epsf,amsfonts,amssymb,epsfig,amsmath,graphics,slashed}
\usepackage{hyperref,graphicx, subfig}
\newcommand{\nn}{\notag \\}

\begin{document}

\makeatletter
\renewcommand{\theequation}{\thesection.\arabic{equation}}
\@addtoreset{equation}{section}
\makeatother

\baselineskip 18pt

\begin{titlepage}

\vfill

\begin{flushright}
\end{flushright}

\vfill

\begin{center}
   \baselineskip=16pt
   {\Large\bf Striped phases from holography}
  \vskip 1.5cm
      Aristomenis Donos\\
   \vskip .6cm
      \begin{small}
      \textit{Blackett Laboratory, 
        Imperial College\\ London, SW7 2AZ, U.K.}
        \end{small}\\*[.6cm]

\end{center}

\vfill

\begin{center}
\textbf{Abstract}
\end{center}

\begin{quote}
We discuss new types of second order phase transitions in holography by constructing striped black holes in $D=4$ with $AdS_{4}$ asymptotics. In the context of $AdS/CFT$, they provide the gravity duals to field theory phases in which translational symmetry is spontaneously broken due to the formation of current density waves. These black holes are associated to three dimensional CFTs at finite temperature and deformed by a uniform chemical potential. We numerically solve a non-linear system of PDEs in order to construct the black hole geometries and extract some of their thermodynamic properties.
\end{quote}

\vfill

\end{titlepage}
\setcounter{equation}{0}


\section{Introduction}

The AdS/CFT correspondence provides a powerful tool for studying strongly coupled systems. Its possible application to problems relevant to condensed matter physics \cite{Hartnoll:2011fn} has attracted a lot of attention over the last few years. One aspect is the phase structure of CFTs at finite temperature and deformed by a chemical potential. Thermal phase transitions of CFTs can be studied in the context of holography via phase transitions of black holes with $AdS$ asymptotics. A canonical example for the bulk dual of the normal phase is the electrically charged AdS-Reissner-Nordstr\"om (AdS-RN) black hole. At low temperatures, these solutions can develop different types of instabilities and the system will be described by a new branch of black hole solutions in general.

A prototype example of such a phase transition is provided by bulk theories which contain scalars that are minimally coupled to the $U(1)$ gauge field used to deform the boundary theory by the chemical potential  \cite{Gubser:2008px}. In this case, when the normal phase becomes unstable against fluctuations of a charged scalar it leads to low temperature black holes where the global $U(1)$ symmetry is broken. The gauge field becomes massive and we are naturally lead to a holographic description of a superconductor/superfluid \cite{Gubser:2008px,Hartnoll:2008vx,Hartnoll:2008kx}.

Another class of instabilities leads to spontaneous breaking of the Euclidian group of symmetries preserved after deforming the theory by a uniform chemical potential. Such phenomena are common in strongly coupled condensed matter systems \cite{vojta} and are worth investigating in the context of holography. The first examples of spatially modulated unstable modes were realized in $D=5$ bulk dimensions in \cite{Ooguri:2010xs,Donos:2011ff} (see also \cite{Domokos:2007kt} for a probe setup). In these five dimensional examples, the unstable modes break translations but they still preserve a Bianchi $VII_{0}$ subgroup giving the dual field theory a helical structure. The slices of constant bulk radius are still homogeneous and the problem of finding the backreacted black holes was reduced to solving a system of ODEs in \cite{Donos:2011ff,Donos:2012wi}. The transition was found to be of second order, in general, while the spatial modulation persists all the way down to zero temperature and the corresponding symmetry breaking ground states were identified in \cite{Donos:2011ff,Donos:2012js}. Similar zero temperature solutions have appeared in a classification of homogeneous branes-like solutions in \cite{Iizuka:2012iv,Iizuka:2012pn}.

Modulated instabilities of electrically charged black holes have also been found in $D=4$ \cite{Donos:2011bh} \footnote{See also \cite{Bergman:2011rf} for a probe brane setup} and more recently in \cite{Donos:2013gda} for any $D\geq 4$. Unlike the homogeneous $D=5$ examples, in this case the backreacted geometry is expected to be inhomogeneous and the solution of PDEs seems necessary. A first construction of such backreacted geometries appeared recently in \cite{Rozali:2012es} for a particular model of the classes studied in \cite{Donos:2011bh}. The authors employed a numerical method developed in \cite{Wiseman:2002zc} to integrate the PDEs resulting from the equations of motion and presented solutions for temperatures below $T\approx 0.9\,T_{c}$. It was found that for a range of temperatures down to $T\approx 0.1\,T_{c}$ the new inhomogeneous solutions where not thermodynamically preferred over the AdS-RN and the system remained dynamically unstable.

Here we will consider two different models, including the one investigated in \cite{Rozali:2012es}, and we will solve the system of PDEs by employing the DeTurck method \cite{Headrick:2009pv,Figueras:2011va} in order to solve Einstein's equations. This method was recently used in \cite{Horowitz:2012ky,Horowitz:2012gs} in order to study transport properties of backgrounds with explicitly broken translations. For both models we are considering, we find a second order transition with the striped phase being thermodynamically favored for all $T<T_{c}$. We present solutions starting from $T\approx 0.999\,T_{c}$ but it is possible to construct solutions arbitrarily close to the critical temperature.

The remaining sections of this paper are organized as follows. In section \ref{sec:the_model} we will introduce the $D=4$ models  in which we will be interested and review the spatially modulated instabilities of the AdS-RN black hole. In section \ref{sec:solutions}, after describing the method we chose to use, we will present the resulting solutions along with their thermodynamic properties. We conclude in section \ref{sec:discussion} with a summary and discussion.

\section{The Einstein-Maxwell-pseudoscalar model}\label{sec:the_model}

We consider the $D=4$ bulk theory which is described by a metric $g_{\mu\nu}$, a gauge field $A_{\mu}$ and a pseudoscalar field $\varphi$
\begin{align}\label{eq:action}
S=\int d^{4}x\sqrt{-g}\,\left(\frac{1}{2}R-\frac{1}{2}\,\left(\partial\varphi\right)^{2}-\frac{\tau(\varphi)}{4}\,F^{2}-V(\varphi)\right)-\frac{1}{2}\int \theta (\varphi)\,F\wedge F
\end{align}
with $F=dA$ being the field strength. The action \eqref{eq:action} yields the equations of motion
\begin{align}\label{eomi}
R_{\mu\nu}-\partial_\mu \varphi\partial_\nu \varphi-g_{\mu\nu}\,V+\tau\,\left(\frac{1}{4}g_{\mu\nu}\,F_{\lambda\rho}F^{\lambda\rho} -F_{\mu\rho}F_{\nu}{}^{\rho}\right)&=0\nn
d\left(\tau\ast F+\vartheta F\right)&=0\nn
d\ast d\varphi+V'*1+\frac{1}{2}\tau'\,F\wedge\ast F+\frac{1}{2}\vartheta'\,F\wedge F&=0\, .
\end{align}
As discussed in \cite{Donos:2011bh}, when the functions $V$, $\tau$ and $\theta$ appearing in the action \eqref{eq:action} admit the expansions
\begin{align}\label{expans}
V=-6+\frac{1}{2}m_{s}^{2}\,\varphi^{2}+\dots,\qquad
\tau=1-\frac{n}{12}\,\varphi^{2}+\dots,\qquad
\vartheta=\frac{c_{1}}{2\sqrt{3}}\,\varphi+\dots\, .
\end{align}
around $\varphi=0$, the equations of motion \eqref{eomi} admit the purely electric AdS-RN black hole solution
\begin{align}\label{eq:RN}
ds_{4}^{2}=&\frac{1}{z^{2}}\,\left(-f(z)\,dt^{2}+\frac{dz^{2}}{f(z)}+dx_{1}^{2}+dx_{2}^{2} \right)\nn
A=&\mu \left(1- z\right),\qquad \varphi=0\nn
f(z)=&\frac{1}{2}\,\left(1-z \right)\,\left(-\mu^{2}z^{3}+4\,z^{2}+4\,z+4 \right)
\end{align}
with $\mu$ being the chemical potential at which the dual field theory is held. In these coordinates, the black hole \eqref{eq:RN} has a regular horizon located at $z=1$ while the $AdS_{4}$ boundary is at $z=0$. A simple surface gravity calculation gives the temperature $T=\left(12-\mu^{2} \right)/\left(8\pi\right)$ showing that the system reaches its zero temperature limit when $\mu=\sqrt{12}$. The near horizon limit of the extremal solution is simply $AdS_{2}\times \mathbb{R}^{2}$ with a non-trivial background gauge field
\begin{align}\label{eq:AdS2limit}
ds_{4}^{2}=&\frac{1}{12}\,ds^{2}\left(AdS_{2}\right)+dx_{1}^{2}+dx_{2}^{2}\nn
F=&\sqrt{12}\,\mathrm{Vol}\left(AdS_{2}\right)
\end{align}

The perturbative analysis of \cite{Donos:2011bh} showed that when the, wavenumber $k$ dependent, matrix
\begin{align}\label{mmfirstmod}
M^{2}&=\left(\begin{array}{ccc}k^{2} & \frac{1}{\sqrt{3}}k & 0 \\24\sqrt{3}k & 24+k^{2}  & -c_{1}k \\0 & -c_{1}k & k^{2}+\tilde{m}_{s}^{2}\end{array}\right), \nn
\tilde{m}_{s}^{2}&=m_{s}^{2}+n
\end{align}
has eigenvalues that violate the $AdS_{2}$ Breitenlohner-Freedman (BF) bound $m^{2}<-3$, the near horizon geometry \eqref{eq:AdS2limit} has tachyonic instabilities. Interestingly, for large enough values of the coupling $c_{1}$, it was found that the lightest tachyonic modes appear at finite values of $k$. In such a  situation, it is natural to expect that there will be a critical temperature, at the onset of the instability, at which a new inhomogeneous branch of black hole solutions will appear. When these solutions exist at lower temperature and are also thermodynamically preferred, the system will undergo a phase transition. These solutions will break the translational symmetries of the normal phase \eqref{eq:RN} down to a discrete group corresponding to the periodicity $2\pi/k$ of the unstable mode. The mode involves the current and momentum densities of the dual field theory and since it is vectorial, parity will be broken at the same time.

Envisaging a continuous transition, its critical temperature can be determined by finding the temperatures $T(k)$ at which the corresponding static modes will appear. The details of these static modes were studied in \cite{Donos:2011bh} and the profile of the temperature as a function of the wavenumber $T(k)$ confirmed that for the cases where the lightest tachyons existed at finite values of $k$, the  maximum of $T(k)$ was also at $k\neq 0$. The critical temperature of the system is precisely at this maximum $T_{c}$.

\begin{figure}
\centering
{\includegraphics[width=8cm]{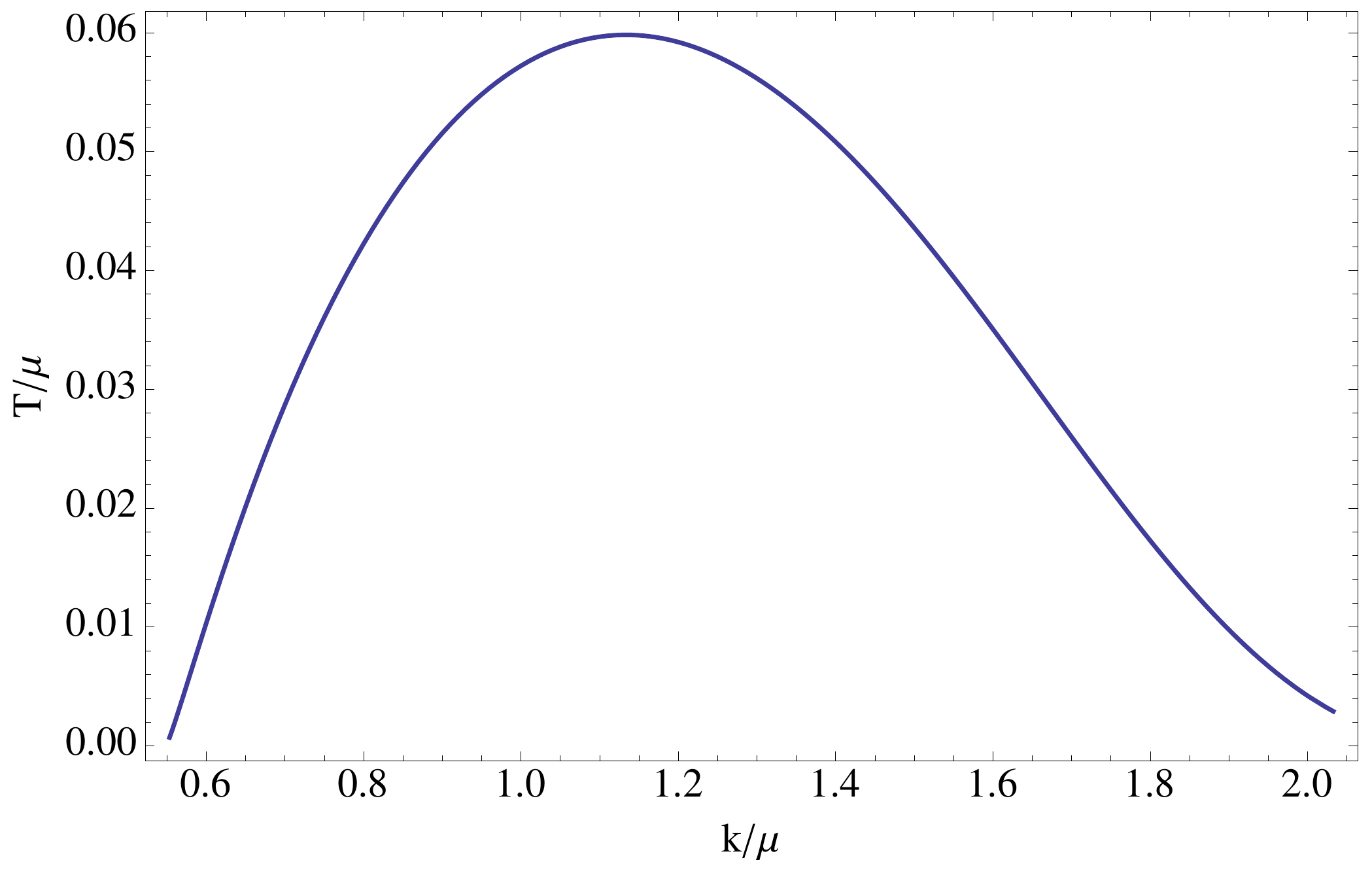}}
\caption{We plot the temperatures $T(k)$ for which we find a static modulated mode as a function of the wavenumber $k$ for the model choice \eqref{eq:vtq_choice} . We find that $k_{c}\approx 1.33\,\mu$ and $T_{c}\approx 0.0598\,\mu$.}
\label{fig:Tk}
\end{figure}

In this paper we will consider two model choices, the first one given by
\begin{align}\label{eq:vtq_choice}
V(\varphi)=-6\,\cosh\left(\sqrt{\frac{2}{3}}\,\varphi \right),\qquad \tau=\frac{1}{\cosh\left(\sqrt{6}\,\varphi \right)},\qquad \theta=\gamma\,\tanh\left( \sqrt{6}\varphi\right)
\end{align}
which has $m_{s}^{2}=-4$, $n=36$ and $c_{1}=6\sqrt{2}\,\gamma$. For values $\gamma>\gamma_{c}=0.9988\ldots$ there exists a wavenumber interval for which  one of the mass matrix \eqref{mmfirstmod} eigenvalues violates the $AdS_{2}$ BF bound $m^{2}(k)<-3$. The model corresponding to the choice $\gamma=1$ is of particular importance since it can be derived as a consistent truncation of $D=11$ supergravity compactified on $SE^{7}$ manifolds \cite{Gauntlett:2009zw,Gauntlett:2009bh,Donos:2012yu}. As the critical value $\gamma_{c}$ is approached, the critical temperature $T_{c}$ gets close to zero and a numerical treatment becomes less accurate. We will take a larger value for $\gamma=1.2$ for which we give a plot of the static mode curve $T(k)$ in figure \ref{fig:Tk}. From this, we deduce that the critical wavenumber is $k_{c}\approx 1.33\,\mu$ while the critical temperature is $T_{c}\approx 0.0598\,\mu$. As we will explain later, the plot in figure \ref{fig:Tk} was made for the quantization choice of the pseudoscalar corresponding to dimension $\Delta=2$.

In order to compare the results of the numerical technique we will use with the findings of \cite{Rozali:2012es}, the second model we will consider has
\begin{align}\label{eq:vtq_choice2}
V(\varphi)=-6-2\,\varphi^{2},\qquad \tau=1,\qquad \theta=\frac{3\,\sqrt{3}}{4}\,\varphi.
\end{align}
 A similar plot with the one in figure \ref{fig:Tk} yields $T_{c}\approx 0.012\, \mu$ and $k_{c}\approx 0.53\,\mu$ \footnote{Our value for $k_{c}$ differs from the one of \cite{Rozali:2012es} by a factor of $\sqrt{2}$ due to the different scale of $x_{1}$ on the boundary. Notice also a typo for the mass of the pseudoscalar $\psi$ in v1 of \cite{Rozali:2012es}.}. In the next section we will focus on these two models and construct the corresponding broken phase black holes for both of them.

\section{Striped black holes}\label{sec:solutions}

\subsection{The setup}

In order to find the backreacted black hole solutions, we need to make a consistent ansatz for the metric and the gauge field. The natural requirement is that it should capture both the normal phase solution \eqref{eq:RN} as well as the perturbative static modes of \cite{Donos:2011bh} when expanded around \eqref{eq:RN}. A suitable ansatz for this purpose is
\begin{align}\label{eq:ansatz}
ds_{4}^{2}&=\frac{1}{z^{2}}\,\left[-f(z)\,Q_{tt}\,dt^{2}+\frac{Q_{zz}}{f(z)}\,dz^{2}+Q_{11}\,\left(dx_{1}+z^{2}Q_{z1}\,dz\right)^{2}+Q_{22}\,\left(dx_{2}+(1-z)\,Q_{t2}\,dt\right)^{2} \right]\nn
A&=\mu\,(1-z)\,a_{t}\,dt+a_{2}\,dx_{2}\nn
\varphi&=z\,h
\end{align}
with all the nine functions appearing in the ansatz $\mathcal{F}=\left\{Q_{tt},\,Q_{zz},\,Q_{ii},\,Q_{z1},\,Q_{t2},\,a_{t},\,a_{2},\,h \right\}$ depending on both $z$ and $x_{1}$ while $f(z)$ is identical with the one that appears in the AdS-RN black hole solution \eqref{eq:RN}. Since we wish to deal with the constraints of Einstein's equations by employing the DeTurck method, we have only partially fixed our coordinates in \eqref{eq:ansatz}. On the other hand, the gauge field ansatz in \eqref{eq:ansatz} leads to  second order equations for $a_{t}$ and $a_{2}$ with no further constrains. This method \cite{Headrick:2009pv,Figueras:2011va} and an appropriate, slightly less general ansatz than \eqref{eq:ansatz}, was recently used in \cite{Horowitz:2012ky,Horowitz:2012gs} in order to explicitly break translations by introducing a lattice deformation of the AdS-RN black hole \eqref{eq:RN}.

Instead of solving Einstein's equations in \eqref{eomi}, we modify it by shifting its left hand side by $-\nabla_{\left(\mu\right.}\xi_{\left.\nu\right)}$ with $\xi^{\mu}=g^{\nu\lambda}\,\left(\Gamma_{\nu\lambda}^{\mu}\left(g\right)-\bar{\Gamma}^{\mu}_{\nu\lambda}\left(\bar{g} \right)\right)$ and $\bar{g}$ being a reference metric which should have the same asymptotic and near horizon structure with the black holes we wish to construct \cite{Headrick:2009pv,Figueras:2011va}. This modification turns Einstein's equations to the so called Einstein - DeTurck equations. For a metric ansatz like our \eqref{eq:ansatz} they are elliptic and after imposing appropriate boundary conditions they should lead to a countable\footnote{When a phase transition happens there are at least two solutions with the same boundary conditions.} set of solutions. We will choose $\bar{g}$ to be the AdS-RN black hole metric which is recovered after setting $Q_{tt}=Q_{zz}=Q_{ii}=1$ and $Q_{z1}=Q_{t1}=0$ in \eqref{eq:ansatz} giving a space-like $\xi$. Our strategy, as in \cite{Horowitz:2012ky,Horowitz:2012gs}, will be to check that the solution generated by solving these equations will have $\xi^{2}\approx 0$ at machine precision. For the solutions that we will present in this work we will have $\xi^{2}<10^{-14}$.

In order to numerically solve the system of the PDEs plus boundary conditions we will discretize the coordinates $z$ and $x_{1}$ on a $N_{z}\times N_{x}$ grid. The aim is to transform the system of differential equations to an algebraic one. In order to approximate the derivatives of our functions at the grid points we will use the pseudospectral collocation method. More specifically we will represent our functions as a finite sum of Chebyshev polynomials for the $z$ coordinate dependence and as a Fourier sum for the periodic $x_{1}$ coordinate. The method we chose to solve the resulting non-linear system of equations is the Newton-Raphson method. Typically, we will be taking $N_{x}=N_{z}=35$ grid points for the first model \eqref{eq:vtq_choice}. For the second model \eqref{eq:vtq_choice2}, we will be mostly interested in getting accurate results for the thermodynamic quantities close to the transition. In that case we will choose $N_{x}=28$ and $N_{z}=70$ while checking that the highest Fourier mode we resolve stays small  down to the temperatures we will construct solutions for.

We now turn to the question of appropriate boundary conditions. In order to avoid the introduction of further deformations to the boundary theory, apart from the chemical potential $\mu$, we have to set the sources of all other operators equal to zero. In particular, for the pseudoscalar $\varphi$ and the spatial component of the gauge field $a_{2}$,  the general fall off near the boundary $z=0$ reads
\begin{align}\label{eq:sc_falloff}
\varphi &=z\,\varphi_{(1)}(x_{1})+ z^{2}\,\varphi_{(2)}(x_{1})+\cdots\nn
a_{2}&=\mu_{j}+j(x_{1})\,z+\mathcal{O}(z^{2})
\end{align}
For the pseudoscalar we may choose between two different quantization conditions corresponding to a dimension $\Delta=1$ or $\Delta=2$ boundary theory operator. We will choose to work with the $\Delta=2$ case in which the function $\varphi_{(2)}$ corresponds to its vev while $\varphi_{(1)}$ corresponds to its source and we should set $\varphi_{(1)}=0$. Imposing the absence of current sources we also require that $\mu_{j}=0$. Under these conditions, the falloff for the remaining fields is
\begin{align}\label{eq:falloff_2}
Q_{tt}&=1+q_{tt}(x_{1})\,z^{3}+\mathcal{O}(z^{4}),\quad Q_{zz}=\mathcal{O}(z^{4}),\quad\nn
Q_{ii}&=1+q_{ii}(x_{1})\,z^{3}+\mathcal{O}(z^{4}),\quad i=1,2\,,\nn
 Q_{t2}&=q_{t2}(x_{1})\,z^{3}+\mathcal{O}(z^{4}),\quad Q_{z1}=\mathcal{O}(z^{2})\nn
 a_{t}&=1-a(x_{1})\,z+\mathcal{O}(z^{2}),\quad
\end{align}
for which the equations of motion also impose
\begin{align}\label{eq:tr_src_free}
q_{tt}(x_{1})+q_{11}(x_{1})+q_{22}(x_{1})&=0\nn
\partial_{x_{1}}q_{11}(x_{1})&=0.
\end{align}
The conditions of equation \eqref{eq:tr_src_free} reflect the fact the energy momentum tensor $T_{\mu\nu}$ of the dual theory, which can be found in appendix \ref{sec:app_thermo}, is traceless  and source free i.e. $\left<T^{\mu}{}_{\mu} \right>=0$ and $\partial_{\mu}\left<T^{\mu}{}_{\nu} \right>=0$.
We are therefore lead to imposing the Dirichlet boundary conditions
\begin{align}\label{eq:AdS4bc}
Q_{tt}(0,x_{1})&=Q_{zz}(0,x_{1})=Q_{11}(0,x_{1})=Q_{22}(0,x_{1})=a_{t}(0,x_{1})=1,\nn
Q_{z1}(0,x_{1})&=Q_{t2}(0,x_{1})=a_{2}(0,x_{1})=h(0,x_{1})=0
\end{align}
on the $AdS_{4}$ boundary.

Regularity of the solution at the horizon, which is at $z=1$, requires that all our functions take finite values and admit an analytic expansion in $(z-1)$
\begin{align}\label{eq:nh_exp}
\mathcal{F}=\mathcal{F}(1,x_{1})+\partial_{z}\mathcal{F}(1,x_{1})\,(z-1)+\cdots
\end{align}
Plugging the near horizon expansion \eqref{eq:nh_exp} in the equations of motion \eqref{eomi} and expanding around $z=1$ we find a condition of constancy of the surface gravity on the horizon $Q_{zz}(1,x_{1})=Q_{tt}(1,x_{1})$ along with eight relations $\mathcal{M}\left(\mathcal{F},\partial_{z}\mathcal{F},\partial_{x_{1}}\mathcal{F},\partial^{2}_{x_{1}}\mathcal{F}\right)=0$ for our horizon data. These nine equations will serve as our boundary conditions on the horizon. It is easy to check that under these conditions the temperature of the black hole will simply be given by $T=\left(12-\mu^{2}\right)/8\pi$.

We also note that both our equations of motion and our boundary conditions are invariant under translations in $x_{1}$, as they should since are are interesting in breaking this symmetry spontaneously. As a result, if $\mathcal{F}(z,x_{1})$ is a solution of our boundary value problem so is $\mathcal{F}(z,x_{1}+b)$ for constant $b$. Since we will be searching for periodic configurations in $x_{1}$ it is natural to expect that within a period, there will be points for which the $\partial_{x_{1}}$ derivatives of our functions will separately vanish. In order to fix this mode we will in addition require that $\partial_{x_{1}}a_{2}(1,0)=0$. In the following section we report on the solutions we find to the posed problem.

\subsection{The solutions}
In order to study the thermodynamics we choose to work in the grand canonical ensemble where the chemical potential $\mu$ is held fixed. In practice, we will be generating solutions by varying $\mu$ in \eqref{eq:ansatz} and by using the scaling symmetry of the boundary theory we will evaluate all the quantities in units of $\mu$.
As previously mentioned, we will be focusing on a particular periodicity in $x_{1}$ for our solutions corresponding to the wavelength $2\pi/k$ of the critical mode with $k=k_{c}\approx 1.33\, \mu$ for e.g. the first model \eqref{eq:vtq_choice}. For temperatures higher than the critical $T_{c}/\mu\approx 0.0598$, the only solution we find to our equations is the RN-black hole \eqref{eq:RN}. In accordance with our expectations, for temperatures $T<T_{c}\approx 0.0598\,\mu$, we find additional solutions for which the functions in our ansatz \eqref{eq:ansatz} have the anticipated periodic profile in $x_{1}$.

\begin{figure}
\vspace{-1cm}
\centering
{\includegraphics[width=7cm]{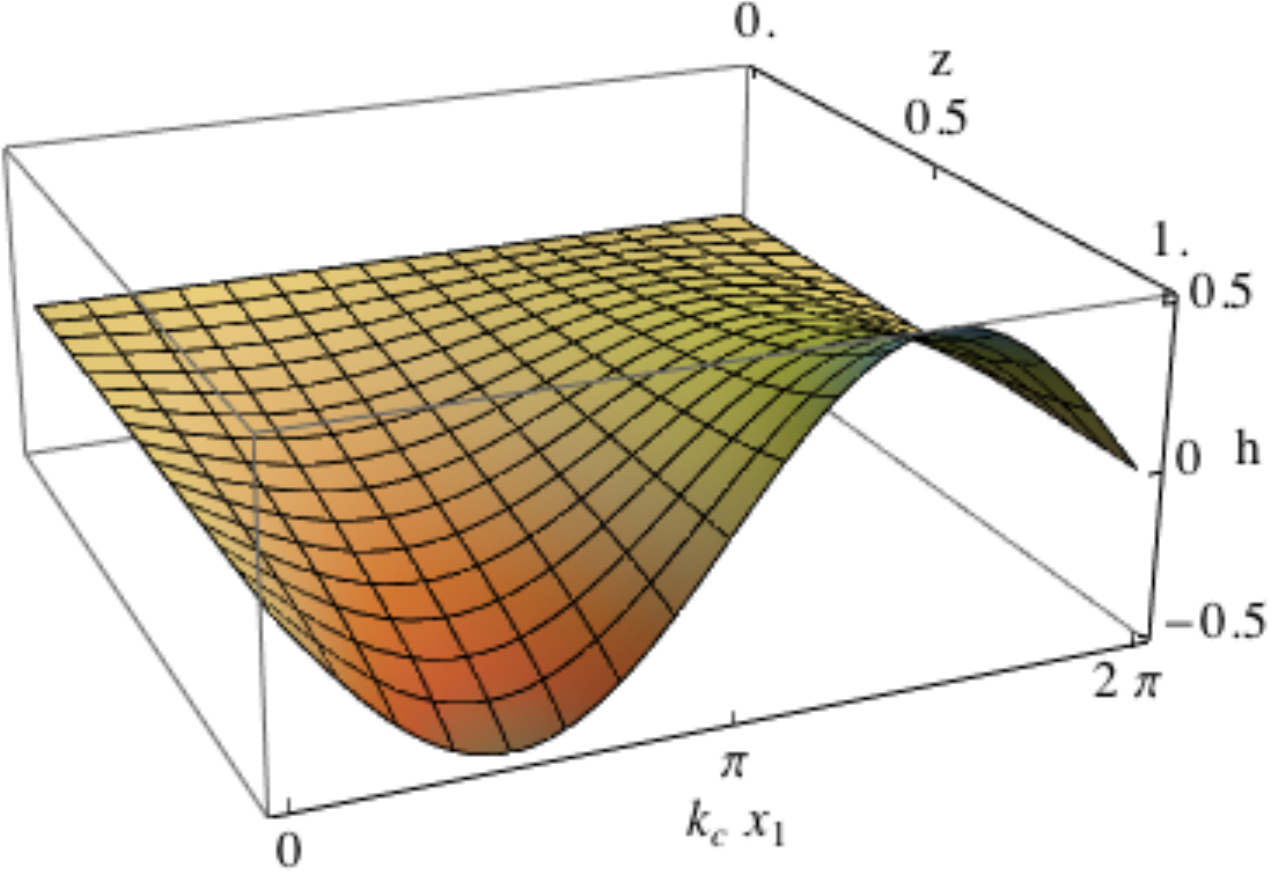}}{\includegraphics[width=7cm]{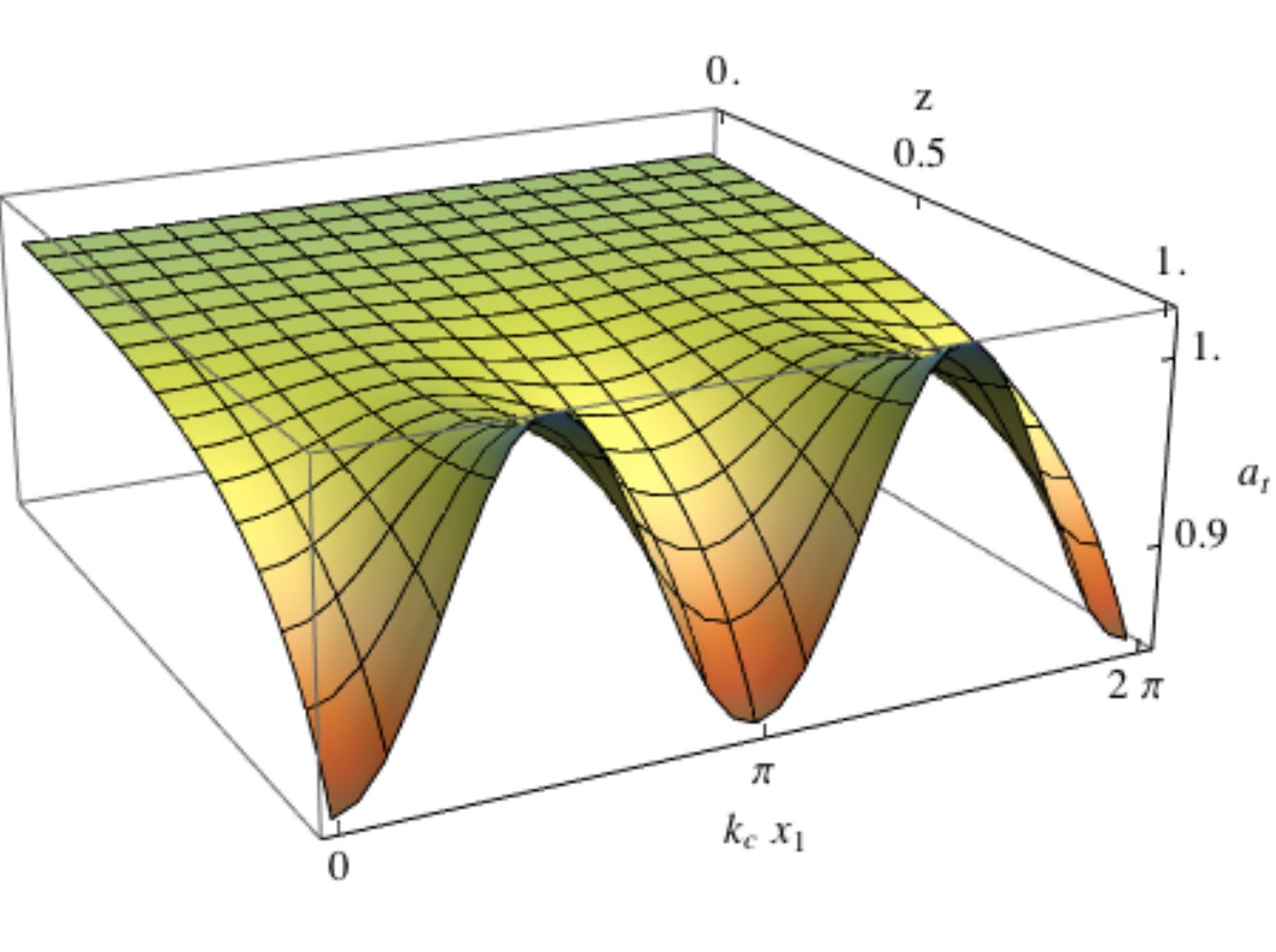}}\\
{\includegraphics[width=7cm]{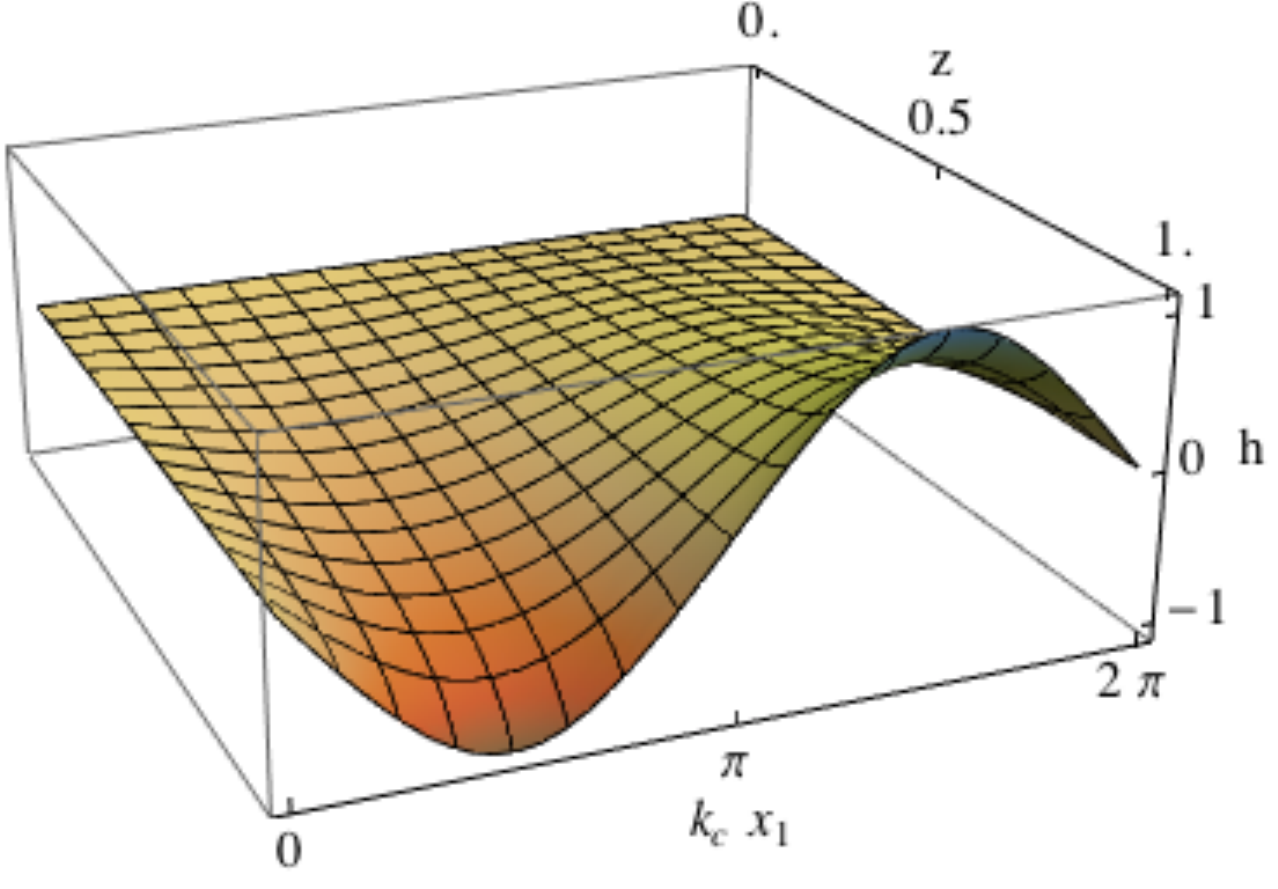}}{\includegraphics[width=7cm]{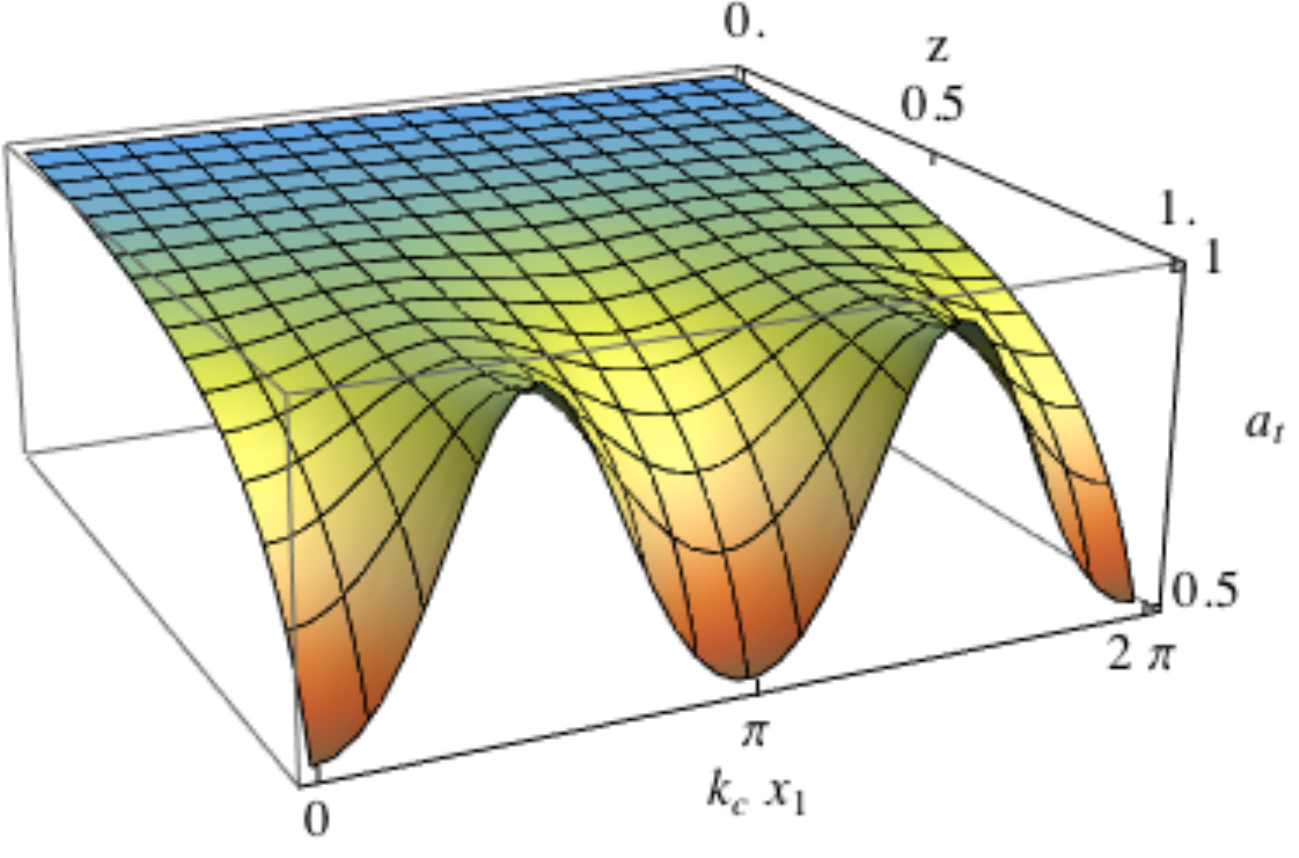}}
\caption{On the left we plot the function $h$ parametrizing the pseudo scalar. On the right we plot the function $a_{t}$. The top plots are for $T=0.6\,T_{c}$, the bottom plots are for $T=0.17\,T_{c}$ and have $N_{x}=40$, $N_{z}=50$.}
\vspace{-0.3cm}
\label{fig:3dfunctions}
\end{figure}

For two such solution, with $T=0.6\,T_{c}$ and $T=0.17\,T_{c}$, we present the profiles of the functions $h$ and $a_{t}$ in figure \ref{fig:3dfunctions} for a single period in $x_{1}$. In \cite{Donos:2011bh} it was shown, using a perturbative series argument, that the electric component of the gauge field $a_{t}$ would be modulated with a period equal to $\pi/k$ close to the transition. For both models \eqref{eq:vtq_choice} and \eqref{eq:vtq_choice2} we chose to study, a similar argument can be made to all orders in perturbation theory in agreement with the periodicity of the right plots in figure \ref{fig:3dfunctions}. We also observe that translation breaking effects become stronger in the IR as we lower the temperature.

The averaged entropy and charge densities of our system can be found through
\begin{align}\label{eq:entropy_FE}
S&=k\,\int_{0}^{2\pi / k}\,\sqrt{Q_{11}(1,x_{1})\,Q_{22}(1,x_{1}) }\,dx_{1}\nn
Q&=\mu+\frac{k\mu}{2\pi}\,\int_{0}^{2\pi / k}\,a(x_{1})\,dx_{1}
\end{align}
while the averaged free energy density is
\begin{align}
\Omega&=M-\mu\,Q-T\,S\nn
M&=2+\frac{\mu^{2}}{2}-\frac{3 k}{2\pi}\,\int_{0}^{2\pi/k}\,q_{tt}(x_{1})\,dx_{1}
\end{align}
with $M$ being the averaged mass density of the black hole, as we explain in Appendix \ref{sec:app_thermo}. In figure \ref{fig:fe_entropy} we give a plot of the averaged entropy density difference between the striped black hole and the RN black hole solution as a function of the temperature. We do the same for the averaged free energy density difference in figure \ref{fig:fe_entropy}. We see that the new phase of black holes leads to lower entropy as well as free energy. More specifically, we find that near the critical temperature $\Delta\left(\frac{\Omega}{\mu^{3}}\right)\propto -\left(1-\frac{T}{T_{c}}\right)^{2}$ and $\Delta\left(\frac{S}{\mu^{2}}\right)\propto \frac{T}{T_{c}}-1$, compatible with a second order phase transition and the first law of thermodynamics\footnote{We keep the period of the solutions in the $x_{1}$ direction fixed.}
\begin{align}
\delta \Omega=-S\,\delta T-Q\,\delta\mu
\end{align}
 for $\delta\mu=0$.
\begin{figure}
\centering
{\includegraphics[width=8cm]{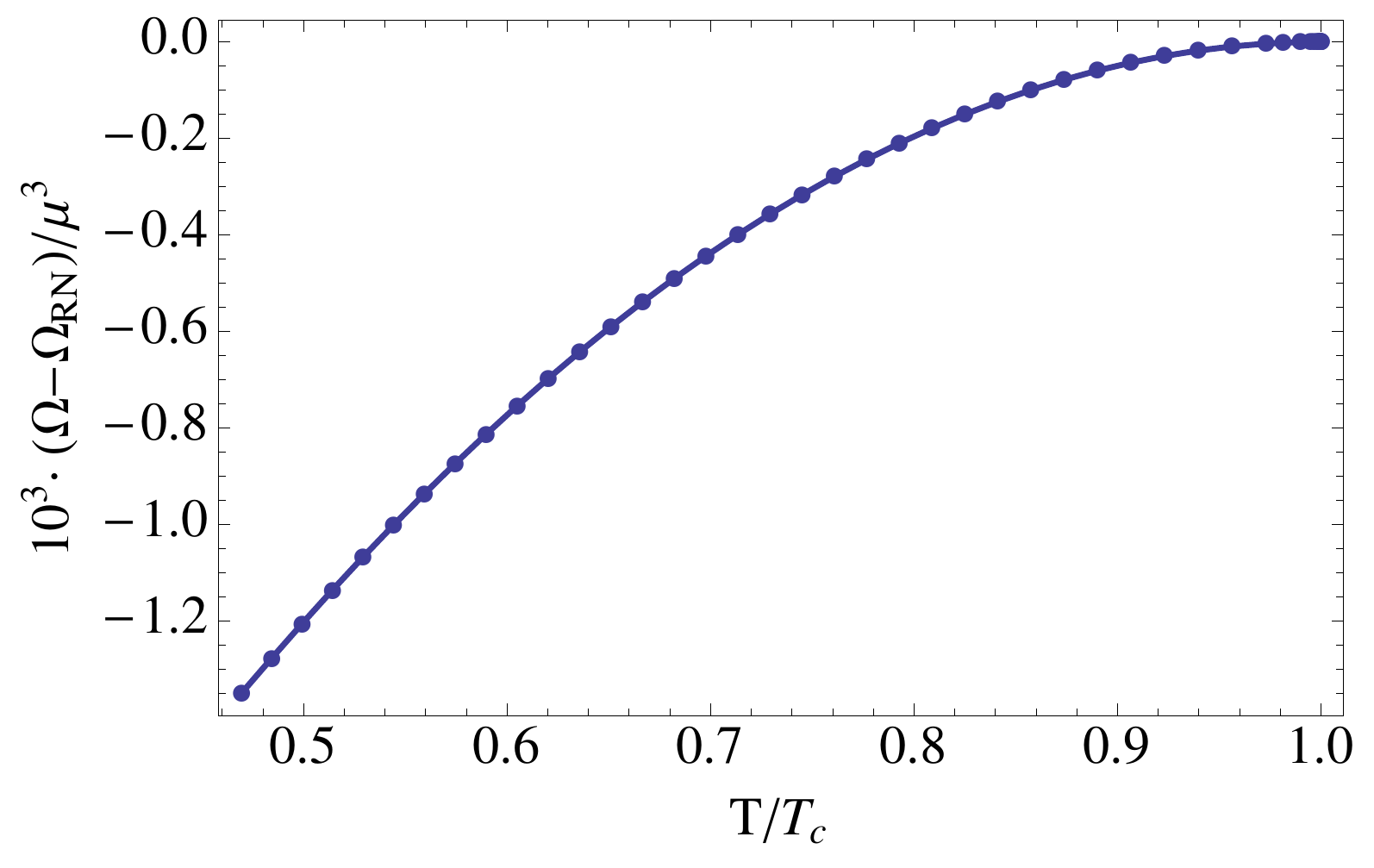}}{\includegraphics[width=8cm]{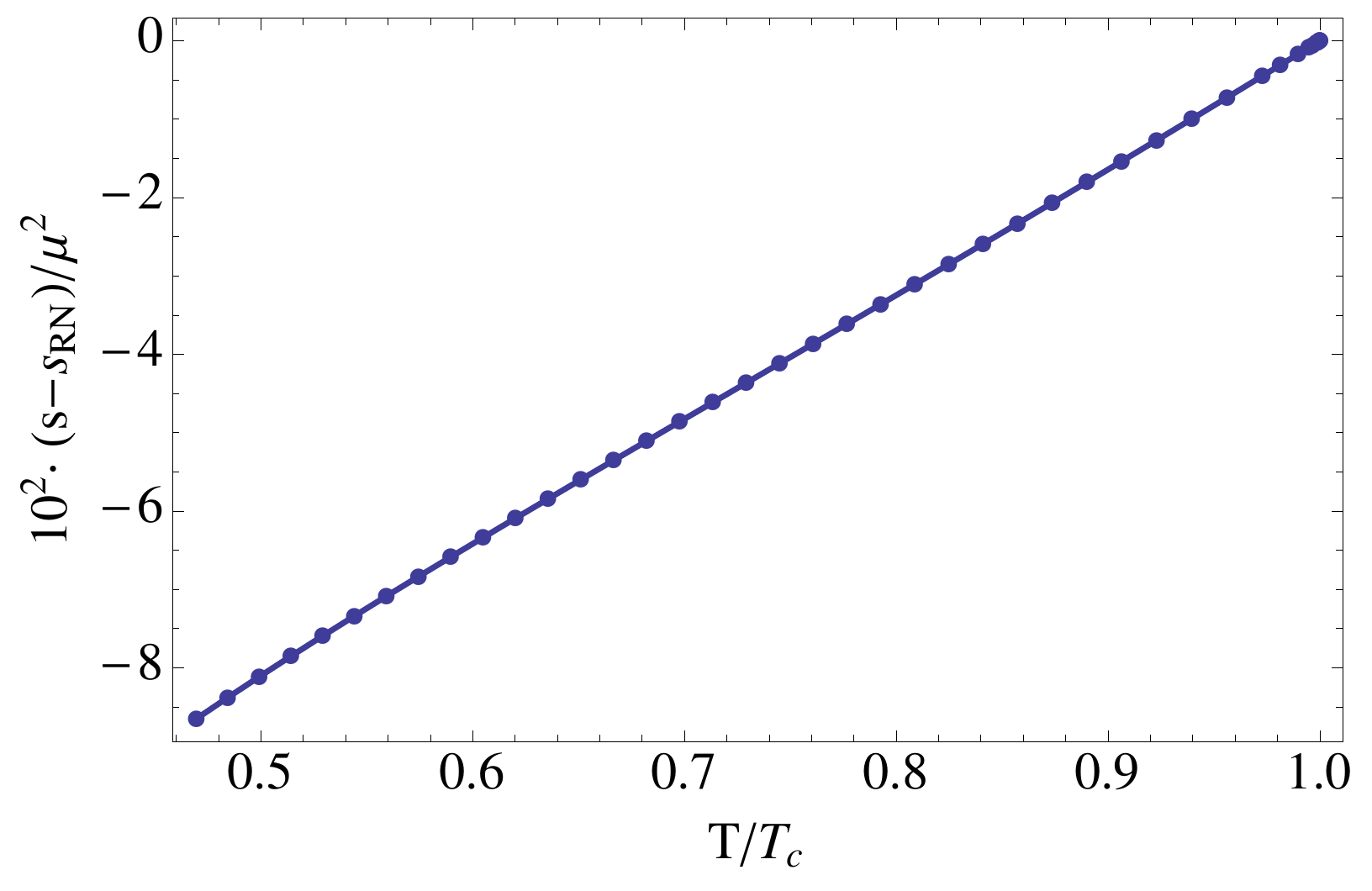}}
\caption{On the left we plot we see that the striped black holes have lower free energy density below the critical temperature $T_{c}$. The right plot shows that the broken phase black holes lead to lower entropy. Both plots are for the model \eqref{eq:vtq_choice}}
\label{fig:fe_entropy}
\end{figure}

\begin{figure}
\centering
{\includegraphics[width=8cm]{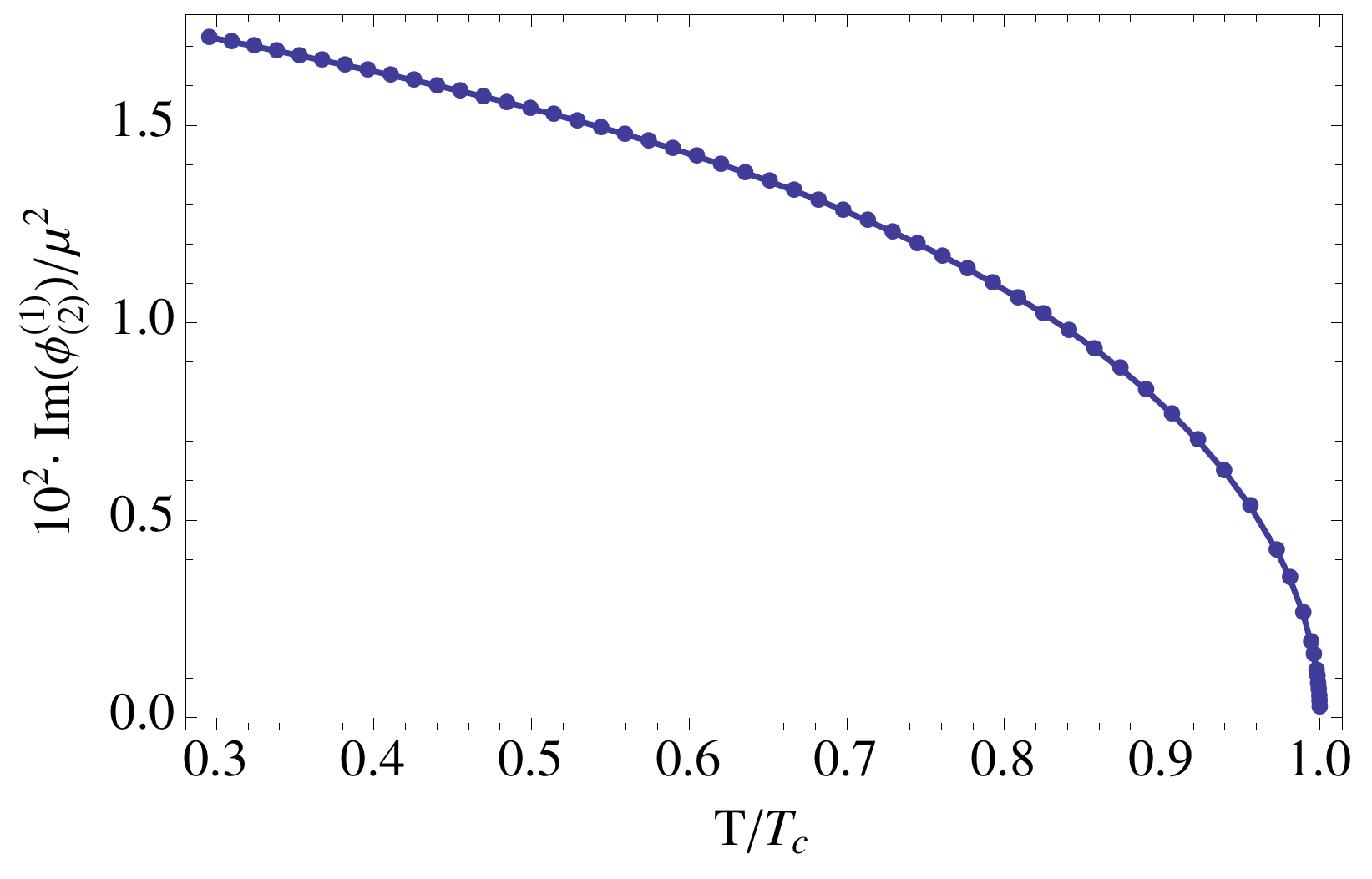}}{\includegraphics[width=8cm]{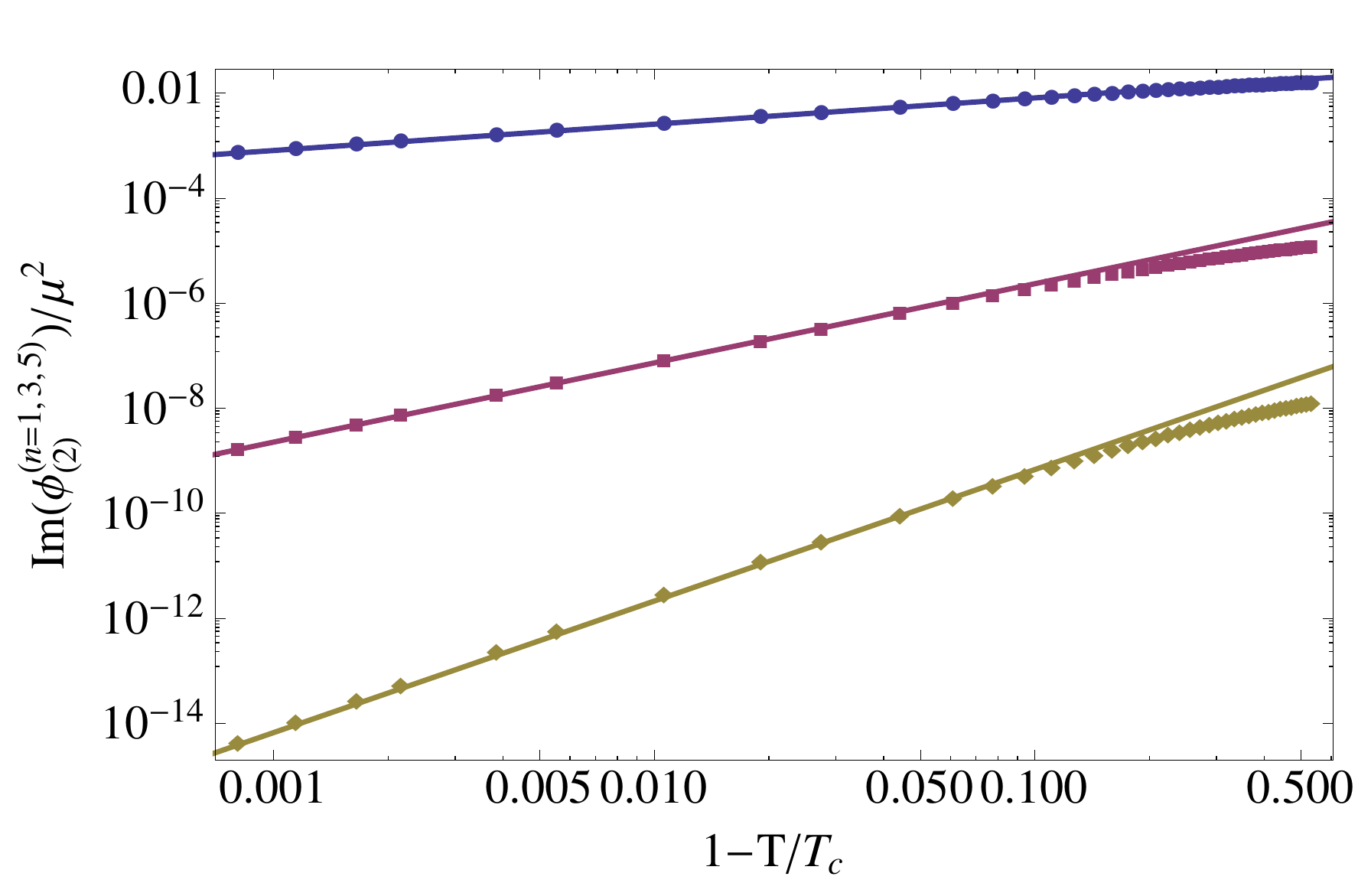}}
\caption{The left plot shows the first non-trivial Fourier mode of the pseudo scalar vev as a function of temperature. The right log-log plot shows its first three non-trivial Fourier modes as a function of $1-T/T_{c}$ along with a linear fit.}
\label{fig:scalarVeV}
\end{figure}

Another set of quantities we will now consider is the Fourier modes of the pseudoscalar operator vev
\begin{align}
\phi_{(2)}^{(n)}=\frac{k}{2\pi}\int_{0}^{2\pi/k}e^{-\imath nkx_{1}}\phi_{(2)}(x_{1})\,dx_{1}.
\end{align}
Close to the critical temperature $T_{c}$, the mode with $n=1$ is expected to have the mean field behavior $\frac{\phi_{(2)}^{(1)}}{\mu^{2}}\propto (1-\frac{T}{T_{c}})^{1/2}$ and the left plot shown in figure \ref{fig:scalarVeV} is compatible with such a power law behavior. In more detail, for the specific model under consideration with even $V(\varphi)$, $\tau(\varphi)$ and odd $\theta(\varphi)$, only modes with odd $n$ will be involved in a perturbative expansion of the pseudoscalar. With the $n$-th mode being sourced at $n$-th order in perturbation theory we expect that we should have the dependence $\frac{\phi_{(2)}^{(n)}}{\mu^{2}}\propto (1-\frac{T}{T_{c}})^{n/2}$. In figure \ref{fig:scalarVeV} we present a log-log plot of $\left|\phi_{(2)}^{(n)}\right|$ as a function of $1-\frac{T}{T_{c}}$ for the first three non-trivial modes corresponding to $n=1,3,5$. A linear fit
\begin{align}
\log\left(\left|\phi_{(2)}^{(n)}\right|\right)=-\alpha_{n}+\gamma_{n}\,\ln\left(1-\frac{T}{T_{c}} \right)
\end{align}
based on our near transition data determines the slopes $\gamma_{1}\approx0.501$, $\gamma_{3}\approx 1.508$ and $\gamma_{5}\approx 2.505$, in good agreement with our previous argument. Similar behavior is found for the momentum $q_{t2}(x_{1})$ and current $j(x_{1})$ density waves.

\begin{figure}
\centering
{\includegraphics[width=8cm]{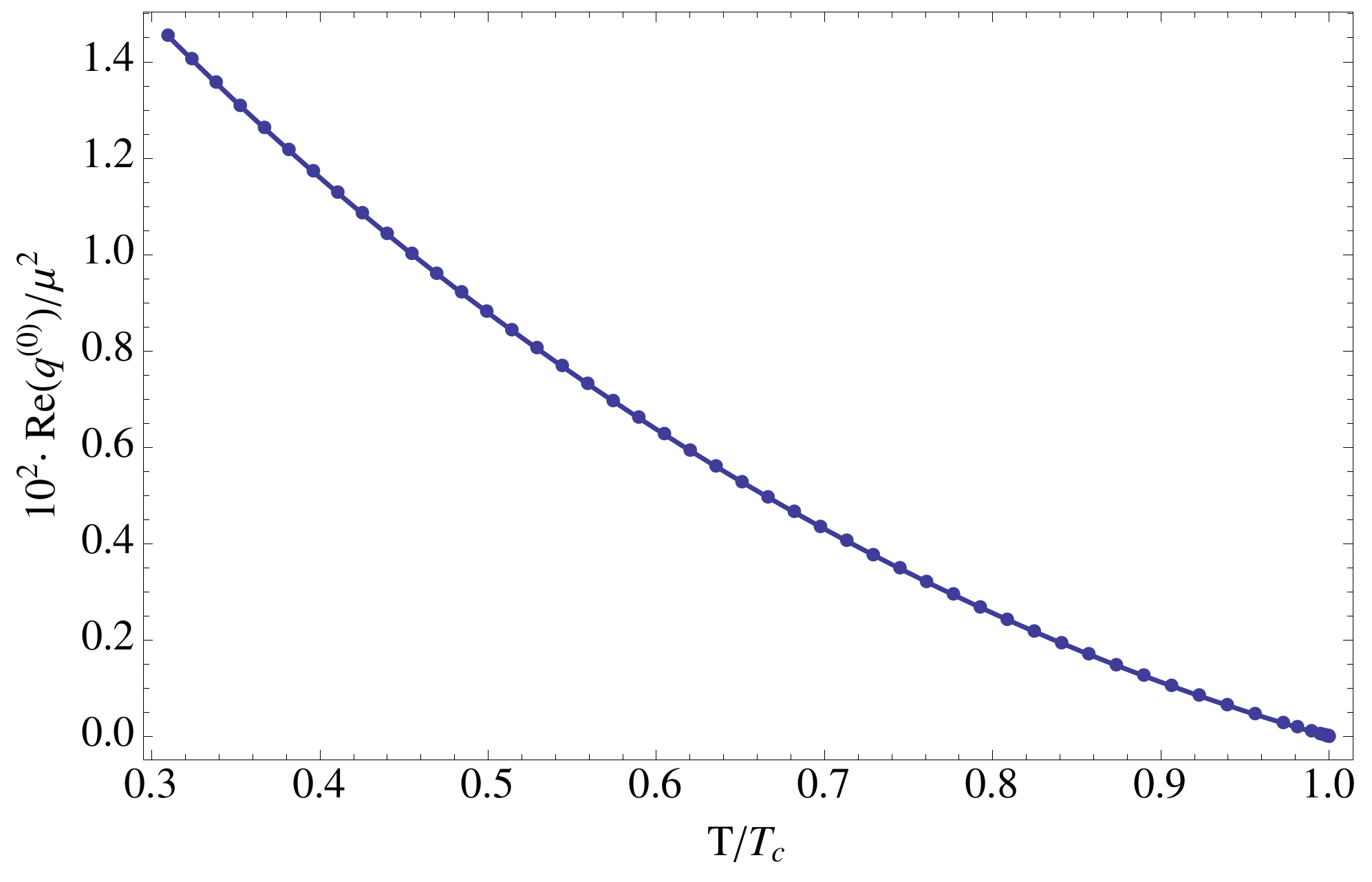}}{\includegraphics[width=8cm]{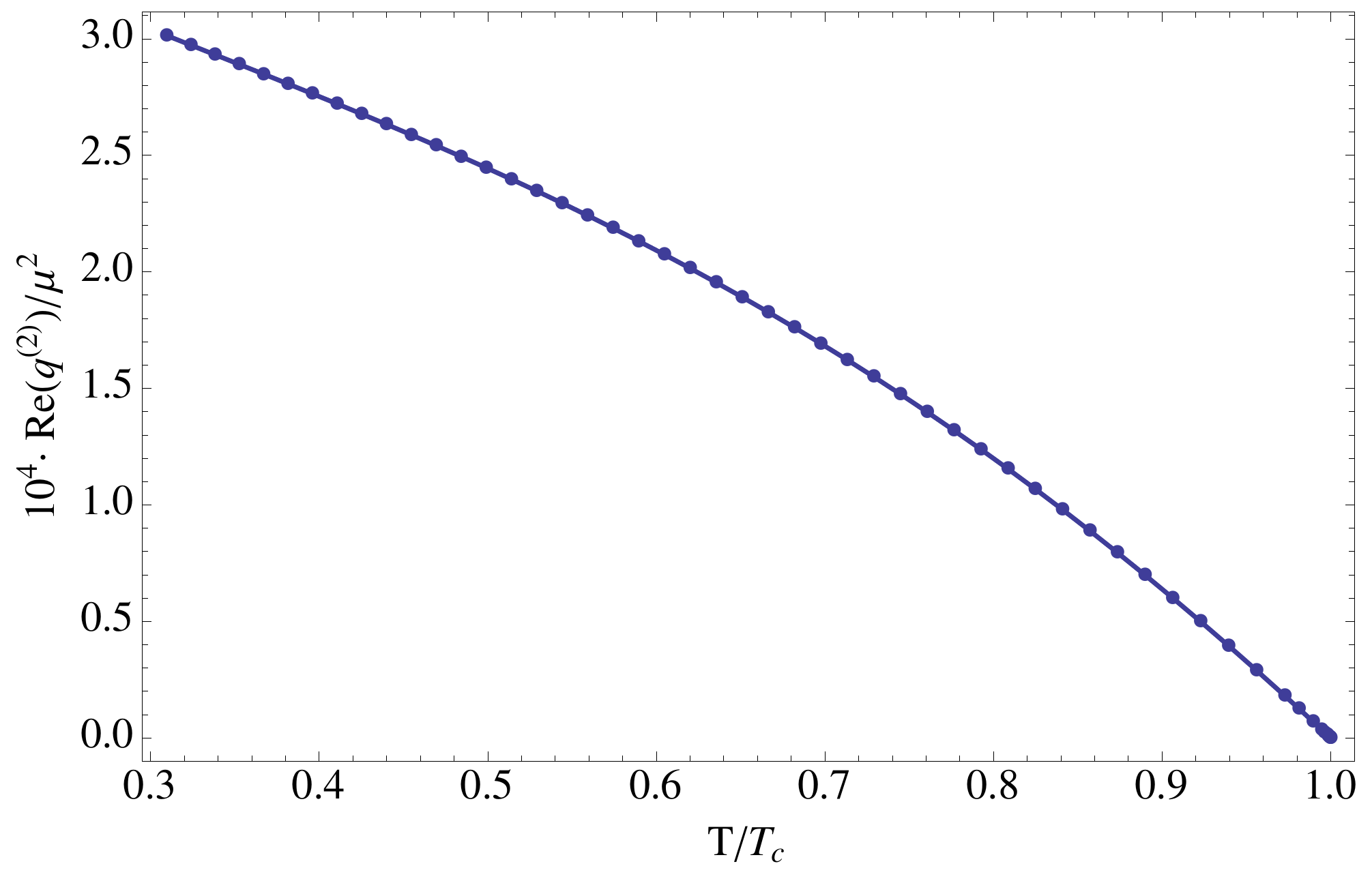}}
\caption{On the left we plot the difference of the average charge density between the striped and the normal phase black hole as a function of temperature. On the right we show the first non-trivial Fourier mode $(n=2)$ of the charge density wave.}
\label{fig:charge}
\end{figure}

We will now examine the charge density wave which forms in our striped phase and whose density profile can be read off from $a(x_{1})$ in \eqref{eq:falloff_2}. In figure \ref{fig:charge} we plot the difference of the averaged charge density $\Delta Q$ from the normal phase as a function of temperature. In the same figure we also plot  its Fourier mode
\begin{align}
q^{(n)}=\frac{\mu k}{2\pi}\,\int_{0}^{2\pi/k}\,e^{-\imath n k x_{1}}a(x_{1})\,dx_{1}
\end{align}
with $n=2$ as a function of temperature. For both cases we observe the expected linear dependence $q^{n=0,2}\propto 1-\frac{T}{T_{c}}$ close to the transition as the electric component of the gauge field $a_{t}$ backreacts at second order.

\begin{figure}
\centering
{\includegraphics[width=8cm]{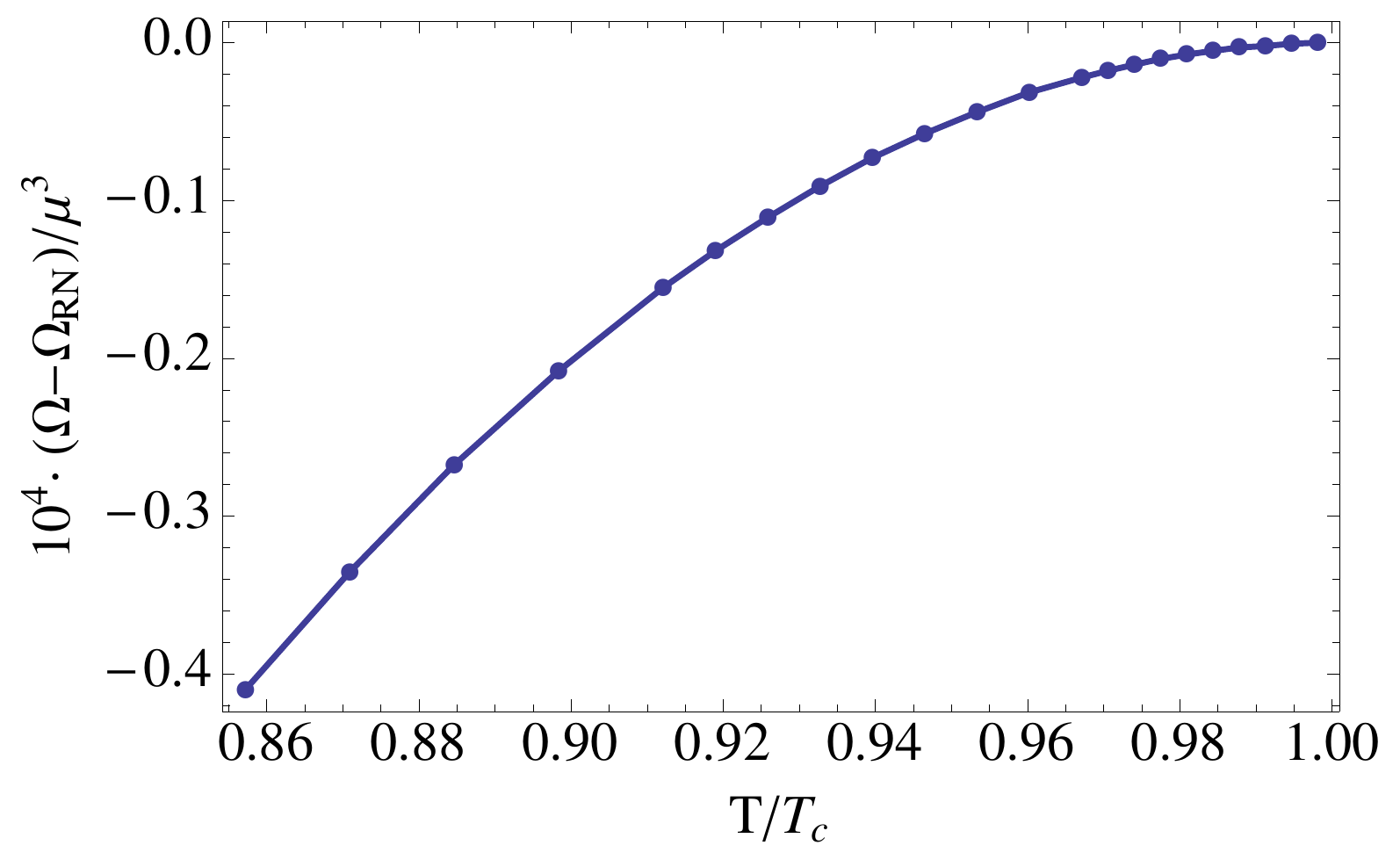}}{\includegraphics[width=8cm]{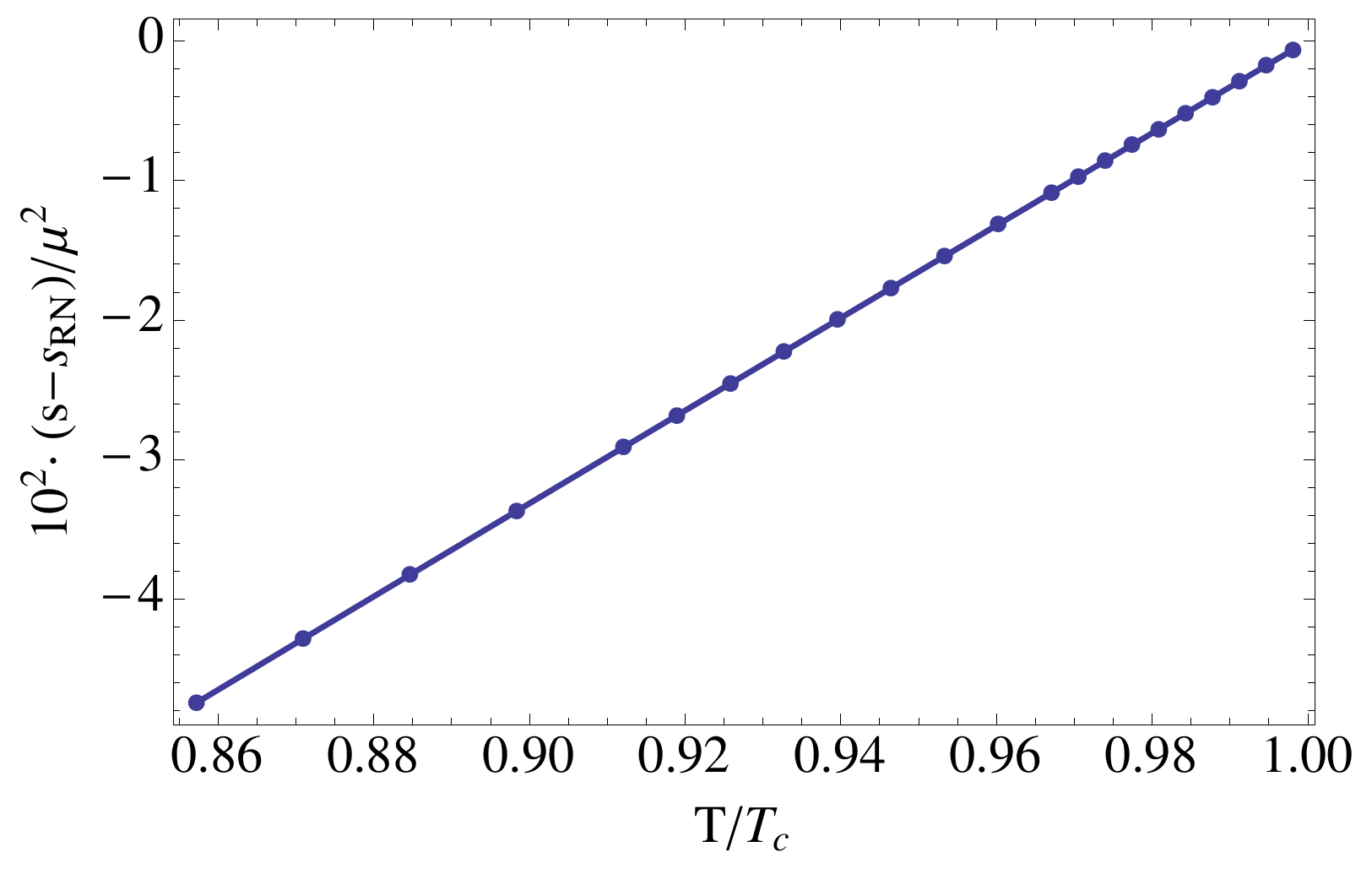}}
\caption{We find similar behavior for the striped black holes of the model \eqref{eq:vtq_choice2}. The striped phase has lower free energy density and entropy below the critical temperature $T_{c}$.}
\label{fig:fe_entropy2}
\end{figure}

We now turn our attention to the model \eqref{eq:vtq_choice2} and repeat the same procedure for $k_{c}\approx 0.53\,\mu$ and $T_{c}\approx 0.012\,\mu$. We don't find any qualitative difference between the two models close to the transition, as we would expect. The transition is again second order and in contrast with the results of \cite{Rozali:2012es}, we find solutions with $T<T_{c}$ which have lower free energy and entropy than the AdS-RN black hole for a given temperature. We  plot the differences of the free energy and entropy densities between the two phases in figure \ref{fig:fe_entropy2}.

In order to calculate the free energy \eqref{eq:entropy_FE} we need to extract the function $q_{tt}(x_{1})$ from the falloff \eqref{eq:falloff_2} of the function $Q_{tt}$ that parametrizes our ansatz \eqref{eq:ansatz} e.g. by taking a third order derivative at $z=0$. Even though the pseudospectral technique we have chosen makes the calculation of derivatives of functions quite accurate, it still is less accurate than the calculation of e.g. the entropy \eqref{eq:entropy_FE} which is using directly the values of the functions. The fact that our data satisfies the first law of thermodynamics at a good accuracy is quite reassuring. In Appendix \ref{sec:alt_param} we outline an equivalent way to phrase the boundary value problem in which the calculation of the normalizable modes of our functions is more direct. We have also solved our PDEs in the way described there and reproduced the plots of figure \ref{fig:fe_entropy2}.

\section{Discussion}\label{sec:discussion}
We numerically constructed four dimensional black hole geometries with inhomogeneous horizons by backreacting on the instabilities of \cite{Donos:2011bh}. These geometries are the bulk duals of CFT phases held at finite temperature and deformed by a uniform chemical potential. The numerical method we chose to follow \cite{Headrick:2009pv,Figueras:2011va} gives results compatible with a second order phase transition. We only reported on the choices \eqref{eq:vtq_choice} and \eqref{eq:vtq_choice2} but since the order of the transition is decided at third order in perturbation theory \cite{Gubser:2001ac} we expect the character of the transition to remain the same for a general class of models. In this work we focused on a particular periodicity for the modulation of our solutions but it is clear from the five dimensional homogeneous analogs \cite{Donos:2012wi,Donos:2012gg} that the configurations of minimum free energy will have a temperature dependent periodicity. The order of the transition is unlikely to change due to this phenomenon in general. An open question is that of the minimum free energy configuration, it is possible that the thermodynamically preferred mode is a superposition in $k-$space. To answer this question one would have to solve PDEs in three dimensions or follow a perturbative approach \cite{Bu:2012mq,Bao:2013fda}.

It would also be interesting to explore the phase transition in theories where the near horizon limit of the normal phase black holes at zero temperature is hyperscaling violating instead of $AdS_{2}\times \mathbb{R}^{2}$. It was recently shown, in \cite{Iizuka:2013ag}, that the near horizon limit of such black holes can have instabilities of the same nature. Based on the general arguments of \cite{Hartnoll:2012pp} the phase transition in those theories is expected to exhibit qualitative differences reflecting the fact that the normal phase black hole branch has zero entropy at its zero temperature limit.

An important question is that of the zero temperature ground state. As we lower the temperature and the non-linearities of the functions $V$, $\tau$ and $\theta$ become important, we expect a rich class of ground states to emerge in the IR. In the absence of modulation and for a purely electric ansatz for the gauge field, the models \eqref{eq:action} reduce to Einstein-Maxwell-Dilaton. In that case, a thorough classification of possible zero temperature IR behavior has been carried out in \cite{Taylor:2008yq,Goldstein:2009cv,Charmousis:2010zz}. It would certainly be interesting to carry out as a similar classification in an inhomogeneous setting. Based on the numerics in \cite{Rozali:2012es}, it was argued by the authors that new IR behavior emerges also in $D=4$. We plan to further explore the low temperature behavior of the striped phases, based on the techniques used in this paper, in the future.

\section*{Acknowledgements}
We would like to thank P. Figueras, J. P. Gauntlett, S. Hartnoll, T. Wiseman and B. Withers for inspiring and valuable discussions.

\appendix

\section{Holographic renormalization}\label{sec:app_thermo}
In order to have a well defined variational problem and a finite total action, we need to supply the action \eqref{eq:action} with appropriate boundary counterterms
\begin{align}
S_{T}=S+\int_{\partial M}\,\sqrt{-\gamma_{\infty}}\,\left( \mathcal{K}-2\sqrt{2}-\frac{1}{\sqrt{2}}\,\varphi^{2}\right)
\end{align}
where the last term ensures that our pseudoscalar has dimension $\Delta=2$. We also have that $\mathcal{K}=\lim_{r\rightarrow0}g^{\mu\nu}\,\nabla_{\mu}n_{\nu}$ with $n^{\mu}$ an outward pointing unit normal vector and $\gamma$ is the determinant of the induced metric on the boundary at $r\rightarrow 0$. Following \cite{Balasubramanian:1999re} and given our asymptotic expansion \eqref{eq:falloff_2} along with the condition \eqref{eq:tr_src_free}, we obtain the expression
\begin{align}
\langle T_{tt}\rangle &=\sqrt{2}\,\left(2+\frac{\mu^{2}}{2}-3\,q_{tt}(x_{1}) \right),\quad \langle T_{x_{i}x_{i}}\rangle=\frac{1}{\sqrt{2}}\,\left(1+\frac{\mu^{2}}{4}+3\,q_{ii}(x_{1}) \right)\nn
\langle T_{tx_{2}}\rangle &=\frac{3}{\sqrt{2}}\,q_{t2}(x_{1})
\end{align}
for the non-trivial component of the boundary stress energy tensor.

The metric of the boundary theory is
\begin{align}
ds_{3}^{2}=\gamma_{(3)}{}_{\mu\nu}dx^{\mu}dx^{\nu}=-2\,dt^{2}+dx_{1}^{2}+dx_{2}^{2}.
\end{align}
After defining the timelike unitary vector $u=2^{-1/2}\,\partial_{t}$, the mass and momentum densities read
\begin{align}
m=&\sqrt{-\gamma_{(3)}}\, u^{\mu}u^{\nu}\,\langle T_{\mu\nu}\rangle=2+\frac{\mu^{2}}{2}-3\,q_{tt}(x_{1})\nn
p_{x_{2}}=&u^{\mu}\langle T_{\mu x_{2}}\rangle=\frac{3}{2}\,q_{t2}(x_{1}).
\end{align}

\section{An equivalent boundary value problem}\label{sec:alt_param}
We will now take advantage of the asymptotic falloff \eqref{eq:falloff_2} to make a better choice of our functions in order to extract the asymptotic data. Instead of using the functions we introduced in the ansatz \eqref{eq:ansatz}, we make the change of variables
\begin{align}\label{eq:redef}
Q_{mm}\left(z,x_{1}\right)&=1+z^{3}\,\hat{q}_{mm}\left(z,x_{1}\right),\quad m=t,\,z,\,1,\,2\nn
Q_{z1}\left(z,x_{1}\right)&=z^{2}\,\hat{q}_{z1}\left(z,x_{1}\right),\quad Q_{t2}\left(z,x_{1}\right)=z^{3}\,\hat{q}_{t2}\left(z,x_{1}\right)\nn
a_{t}\left(z,x_{1}\right)&=1-z\,\hat{a}\left(z,x_{1}\right),\quad a_{2}\left(z,x_{1}\right)=z\,\hat{j}\left(z,x_{1}\right)\nn
\varphi&=z^{2}\,\hat{\varphi}_{(2)}\left(z,x_{1}\right).
\end{align}
The aim is to solve the problem in such a way that the values of the functions we are solving for are directly related to the values of the vevs we wish to calculate. Our requirement for a regular horizon at $z=1$ leads us again to imposing very similar boundary conditions with the ones described in section \ref{sec:solutions} for the functions $\hat{\mathcal{F}}=\left\{ \hat{q}_{mm},\,\hat{q}_{rx_{1}},\,\hat{q}_{tx_{2}},\,\hat{a},\,\hat{j},\,\hat{\varphi}_{(2)}\right\}$. For example we find that we should impose the zeroth law of thermodynamics guaranteed by $\hat{q}_{tt}\left(1,x_{1}\right)=\hat{q}_{rr}\left(1,x_{1}\right)$.

The difference in the nature of the boundary conditions we need to impose comes from the $AdS_{4}$ boundary at $z=0$. Plugging the ansatz in the equations of motion \eqref{eomi} we find the simple first order relations at $z=0$
\begin{align}\label{eq:ads4_bc_2}
\partial_{z}\hat{q}_{tt}&=-\hat{\varphi}_{(2)}^{2}+\frac{1}{4}\mu^{2}\,\hat{a}\,\left(\hat{a}+2 \right),\quad 
\partial_{z}\hat{q}_{rr}=\frac{1}{2}\,\hat{j}^{2}+3\,\hat{\varphi}_{(2)}^{2} -\frac{1}{4}\mu^{2}\,\hat{a}\,\left(\hat{a}+2 \right)\nn
\partial_{z}\hat{q}_{11}&=-\hat{\varphi}_{(2)}^{2},\quad 
\partial_{z}\hat{q}_{22}=-\frac{1}{2}\,\hat{j}^{2}-\hat{\varphi}_{(2)}^{2},\quad
\partial_{z}\hat{q}_{z1}=\hat{\varphi}_{(2)}\,\partial_{x_{1}}\hat{\varphi}_{(2)}\nn
\partial_{z}\hat{q}_{t2}&=\hat{q}_{t2}+\frac{1}{2}\mu\,\hat{j}\,\left(\hat{a}+1\right),\quad
\partial_{z}\hat{a}=\hat{a},\quad
\partial_{z}\hat{j}=0,\quad
\partial_{z}\hat{\varphi}_{(2)}=0
\end{align}
which we can use as our boundary conditions. We also find that that we should satisfy the conditions
\begin{align}\label{eq:ads4_const}
\hat{q}_{tt}\left(0,x_{1}\right)+\hat{q}_{x_{1}x_{1}}\left(0,x_{1}\right)+\hat{q}_{x_{2}x_{2}}\left(0,x_{1}\right)=0,\quad \partial_{x_{1}}\hat{q}_{x_{1}x_{1}}\left(0,x_{1}\right)=0,\quad \hat{q}_{zz}\left(0,x_{1}\right)=0
\end{align}
Given the above boundary conditions for the functions $\hat{\mathcal{F}}$, we can reproduce the plots of figure \ref{fig:fe_entropy2} at relatively low grid resolutions e.g. $N_{x}=25$ and $N_{z}=45$. We also check that the resulting solutions satisfy the constraints \eqref{eq:ads4_const}.

\bibliographystyle{utphys}
\bibliography{helical}{}
\end{document}